\documentclass[sigconf,nonacm]{acmart}
\usepackage{amsmath,booktabs,graphicx}
\usepackage{pgfplots}
\usepgfplotslibrary{groupplots}
\pgfplotsset{compat=1.18}
\usepackage{xurl}
\graphicspath{{figures/}}
\hypersetup{hidelinks,keeppdfinfo,pdfauthor={Shengqi Zhou, Zhigang Liu, Wenxiang Zhu, Boyong Du, Kairui Yang, Haicheng Huang, Weijian Bu}}
\setcopyright{none}
\renewcommand\footnotetextcopyrightpermission[1]{}
\acmDOI{}\acmISBN{}
\newcommand{\admissible}{\mbox{\textsc{Admissible}}}
\newcommand{\reject}{\mbox{\textsc{Reject}}}
\newcommand{\probe}{Joint probe}

\begin{document}
\title[When Inputs Receive No Score]{When Inputs Receive No Score: Evaluating Open-Web Event Admission on the Complete Stream}

\author{Shengqi Zhou}
\authornote{These authors contributed equally.}
\authornote{Corresponding author.}
\email{shbiao@stu.nepu.edu.cn}
\affiliation{%
  \institution{Northeast Petroleum University}
  \city{Daqing}
  \state{Heilongjiang}
  \postcode{163318}
  \country{China}}

\author{Zhigang Liu}
\authornotemark[1]
\affiliation{%
  \institution{Northeast Petroleum University}
  \city{Daqing}
  \state{Heilongjiang}
  \postcode{163318}
  \country{China}}

\author{Wenxiang Zhu}
\authornotemark[1]
\affiliation{%
  \institution{Tongji University}
  \city{Shanghai}
  \postcode{200092}
  \country{China}}

\author{Boyong Du}
\affiliation{%
  \institution{Chongqing University}
  \city{Chongqing}
  \postcode{400044}
  \country{China}}

\author{Kairui Yang}
\affiliation{%
  \institution{Northeast Petroleum University}
  \city{Daqing}
  \state{Heilongjiang}
  \postcode{163318}
  \country{China}}

\author{Haicheng Huang}
\affiliation{%
  \institution{Northeast Petroleum University}
  \city{Daqing}
  \state{Heilongjiang}
  \postcode{163318}
  \country{China}}

\author{Weijian Bu}
\affiliation{%
  \institution{Northeast Petroleum University}
  \city{Daqing}
  \state{Heilongjiang}
  \postcode{163318}
  \country{China}}

\renewcommand{\shortauthors}{Zhou et al.}

\begin{abstract}
A Web item can link to the correct event without contributing information worth organizing. Evaluating item-level admission therefore begins with complete inputs, including those a processing pipeline fails to score. We study how this upstream selection changes ranking comparisons and constrains operational decisions. Across six instruction-model backbones, three strategies, and 602 human-annotated evaluation items, systems score different populations with different class compositions. Comparing rankings on shared inputs separates these population differences from discrimination. Score availability also limits attainable retention: four configurations score only ten of twelve positive development items, making a 90\% full-stream target unreachable by threshold adjustment. Paired precision experiments show that this constraint depends on execution configuration as well as model choice. Qwen scores the same number of evaluation items under INT4 and INT8, yet exchanges four scored items for four others. Extraction-recovery diagnostics further show that obtaining more scores can increase false acceptance under an unchanged threshold. We connect these observations through a complete-stream evaluation protocol that checks calibration feasibility and separates upstream score loss from decision loss. The resulting account explains which inputs a system can score, what its operating point can achieve, and what it actually admits.
\end{abstract}


\maketitle
\hypersetup{pdftitle={When Inputs Receive No Score: Evaluating Open-Web Event Admission on the Complete Stream},pdfauthor={Shengqi Zhou, Zhigang Liu, Wenxiang Zhu, Boyong Du, Kairui Yang, Haicheng Huang, Weijian Bu}}

\section{Introduction}
\label{sec:intro}
\begin{figure*}[t]
\centering
\includegraphics[width=.82\textwidth]{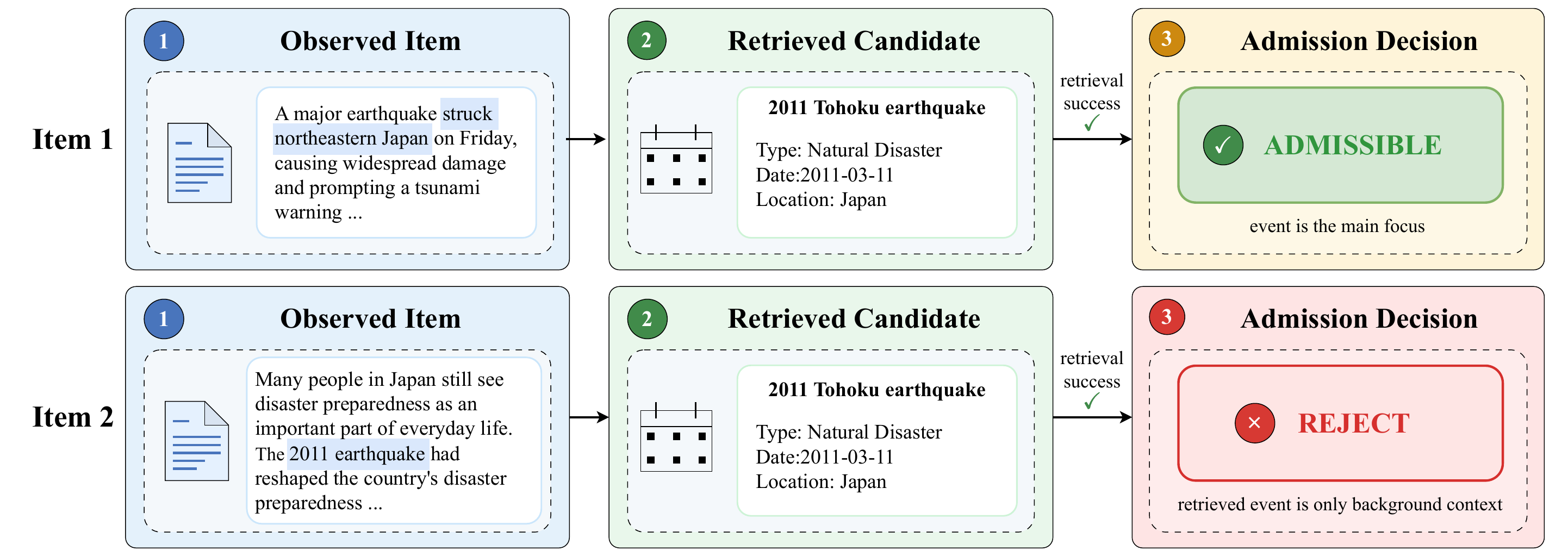}
\caption{Successful retrieval does not determine admission. Two Web items retrieve the same earthquake, but the event is central to the news report and background context in the preparedness post. Admission concerns the complete item.}
\Description{Two Web items retrieve the same earthquake candidate. The current-event report is admitted, while the preparedness post using the earthquake as background is rejected.}
\label{fig:phenomenon}
\end{figure*}

A Web post can mention an event without being about that event. A news report on an earthquake can contribute an event record, while a preparedness tip citing the same earthquake contributes advice. Both may retrieve the correct knowledge-base entry. The decision to organize the complete item remains different (Figure~\ref{fig:phenomenon}). Conversely, an emerging event can warrant organization before any corresponding entry exists. Before asking which event an item refers to, we first ask whether the item should enter the event-organization pipeline at all. We call this decision \emph{item-level event admission}. In pipelines that use large language models for extraction and candidate assessment, processing non-admissible items still consumes retrieval and inference resources. Better linking alone does not resolve which items warrant that processing. Evaluating this admission boundary is therefore part of assessing how open-Web information is organized.

Event-linking evaluation starts from selected mentions and asks which target each refers to, including NIL when no target is available \citep{nothman2012event,eventlinking2023,argumentaware2024}. Admission starts earlier, with the incoming Web item. This change of unit exposes a measurement problem: a pipeline must first extract candidates or construct features before it can produce an admission score. A direct strategy can score the item without those steps. Given the same 100 inputs, one system might score all 100 while another scores only 80. Their AUCs summarize different populations. The missing 20 items disappear from the second ranking metric but remain part of the stream the system must process.

This selection also limits the decisions a system can make. Suppose only ten of twelve qualifying development items receive scores. Even admitting all ten cannot meet a 90\% full-stream retention target. Lowering the threshold affects scored items, not the two lost upstream. Conversely, recovering their scores exposes additional inputs to the decision rule, including inputs that should be rejected. Better upstream processing can therefore change operational errors without improving the ranking metric or selecting a suitable threshold. For Web monitoring and event organization, the relevant outcome is what enters the organized stream, not just how successfully scored items are ordered.

We evaluate these dependencies together. The scored population is treated as a system outcome. Ranking comparisons use identical scored subsets, calibration first checks whether enough qualifying items can receive scores, and operational rates retain all incoming items in their denominators. A loss decomposition locates qualifying items lost before scoring separately from those rejected at the chosen threshold. These measurements connect an upstream engineering change to its downstream consequences.

Two complementary experiments establish the importance of this distinction. Six-backbone evaluation shows that candidate-dependent and direct strategies reach different populations, often excluding different proportions of the two Gold classes. Paired inference-precision experiments then hold the backbone and input IDs fixed. Even unchanged aggregate coverage can hide an exchange of scored items, and a precision change can make a development target feasible while evaluation coverage declines. Shared-subset comparisons, threshold sensitivity, and extraction recovery explain how these changes affect the interpretation of system performance.

Our contributions are:
\begin{itemize}
\item \textbf{A complete-stream evaluation framework for event admission.} We separate item eligibility from target availability and connect the scored population to calibration feasibility, shared-subset ranking, and upstream versus decision loss.
\item \textbf{Evidence that score availability is an operational system property.} Six-backbone comparisons and paired precision perturbations show how processing choices change the evaluated population and attainable operating points. Recovery and fixed-score calibration analyses isolate their consequences for complete-stream decisions.
\item \textbf{A layered resource supporting reproducible evaluation.} The 3,290-record resource separates human evaluation and calibration from AI-assisted observation. Item-level scores, availability masks, annotation records, and analysis code support the comparisons without evaluation-set tuning.
\end{itemize}

\section{Related Work}
\label{sec:related}

\paragraph{Event linking and item admission.}
Event linking grounds selected mentions to news archives or knowledge-base events \citep{nothman2012event,eventlinking2023}, with event arguments helping disambiguation \citep{argumentaware2024}. Entity-linking research addresses incomplete inventories, NIL prediction, and language-model disambiguation \citep{mcnamee2010kbp,learnnotlink2023,yuan2023partial,ye2025llm}. These tasks establish a mention's target or its absence. Admission concerns the role of an event in the complete item: a linkable historical reference can be background, while a new event can be central without a known target. Target availability and item eligibility therefore define different decisions.

\paragraph{Detection, salience, and organization.}
Web event detection ranges from real-time social sensing \citep{sakaki2010earthquake} to graph-based detection \citep{qiu2024metapath} and retrieval-augmented event discovery and evolution \citep{liu2026rag}. Salience identifies events carrying a document's central information \citep{liu2018salience,jindal2020salient}. Admission additionally asks whether that information establishes an organizing target: a salient hypothetical can be rejected, while a concrete reported action can qualify without being globally important. Timeline construction, background summarization, and cross-document coreference organize or connect event information \citep{li2021timeline,pratapa2023background,chen2025coreference,chen2025discourse}. We examine the item-level boundary at which information enters such workflows.

\paragraph{Selective prediction and calibration.}
Reject-option and selective-prediction methods study predictive risk as a system withholds decisions \citep{chow1970reject,elyaniv2010selective,geifman2019selectivenet}. Our distinction is between a threshold rejecting an available score and upstream processing preventing that score from existing. Probability calibration concerns the reliability of predicted probabilities \citep{guo2017calibration}. We instead study development-selected operating thresholds, including targets that unavailable scores make infeasible. Work on unanswerable RAG queries examines related output decisions \citep{peng2025unanswerability}. Tracking pre-score losses makes the inputs unavailable to the final decision rule explicit.

\paragraph{Evaluation populations.}
Benchmark construction, retrieval-based sampling, and impression logs influence which examples are evaluated \citep{vanerp2016evaluating,born2020dataset,ahn2024impression}. Full-stage ranking addresses selection across retrieval and ranking stages \citep{zheng2024fullstage}, while recent information-extraction work examines sampling and evaluation sensitivity \citep{huang2026reevaluating}. Here, selection also arises during inference on a fixed incoming cohort. The same inputs yield different scored subsets across systems and execution conditions. We connect this population change to both the meaning of ranking comparisons and the feasibility of complete-stream operating targets.

\begin{figure*}[t]
\centering
\includegraphics[width=\textwidth]{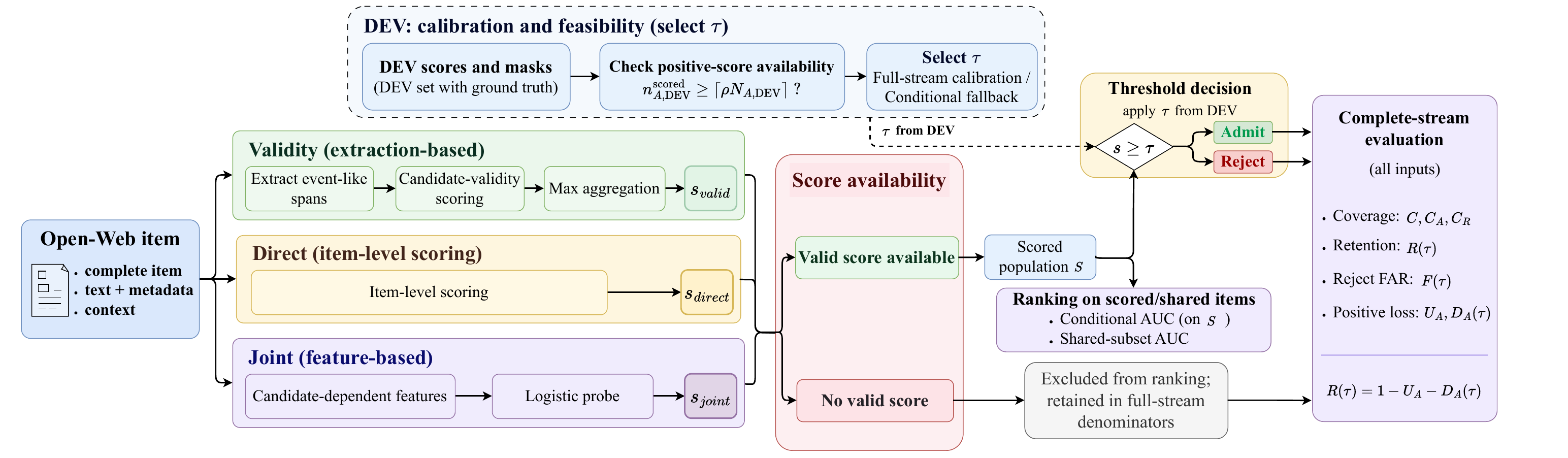}
\caption{Scoring and evaluation pipeline, applied separately to each strategy. Validity uses extraction without target retrieval, Direct scores complete items, and Joint combines candidate-dependent features. DEV scores and availability masks determine calibration feasibility and the threshold. Ranking uses scored or shared items, while complete-stream evaluation retains both scored and unscored inputs.}
\Description{Three independent scoring strategies produce strategy-specific scores and availability masks. Scored items enter ranking evaluation and threshold decisions. Unscored items bypass both but remain in complete-stream evaluation. A separate development calibration branch supplies the decision threshold.}
\label{fig:pipeline}
\end{figure*}

\section{Admission and Complete-Stream Evaluation}
\label{sec:contract}

\subsection{The item and its score}

An input $x_i$ is a complete Web item with its displayed text and recorded context. Its Gold label is \admissible{} ($y_i=1$) when the evidence foregrounds an identifiable real-world occurrence, action, change, result, or institutional act suitable for organization. Otherwise it is \reject{} ($y_i=0$). Historical and personal events can qualify when central to the item. A hypothetical or a background reference does not qualify on that basis alone. Annotation uses the displayed body, timestamp, and reply, quote, and media-presence indicators, without opening links or independently checking factual truth. A known target, exact date, or named location is not required.

A candidate-dependent route extracts event-like expressions and may retrieve targets before constructing a score. Direct admission scores the item without extraction or retrieval, although direct scoring can also fail. Extraction deliberately includes expressions broader than the admission policy, including hypothetical and background references. Finding an expression therefore differs from deciding that the item qualifies. Figure~\ref{fig:pipeline} summarizes the scoring routes, DEV calibration, and evaluation populations.

For a fixed system and cohort, let $c_i=1$ when item $i$ receives a valid score $s_i$, and $c_i=0$ otherwise. The scored population is $\mathcal{S}=\{i:c_i=1\}$. Larger scores favor admission. The evaluated systems admit an item when it is scoreable and $s_i\geq\tau$, giving action $a_i(\tau)=1$ in that case and zero otherwise. An unscored item retains its Gold label and remains in the evaluation denominator.

\subsection{Where qualifying inputs are lost}

Let $N_A=\sum_i y_i$ and $N_R=\sum_i(1-y_i)$. We measure overall and class-conditional score coverage, full-stream retention, and Reject false acceptance as
\begin{equation}
\begin{aligned}
C&=\frac{\sum_i c_i}{N},&
C_A&=\frac{\sum_i y_i c_i}{N_A},\\
C_R&=\frac{\sum_i(1-y_i)c_i}{N_R},&
R(\tau)&=\frac{\sum_i y_i a_i(\tau)}{N_A},\\
F(\tau)&=\frac{\sum_i(1-y_i)a_i(\tau)}{N_R}.
\end{aligned}
\label{eq:rates}
\end{equation}
$F$ is the fraction of all Gold-Reject inputs admitted, not their fraction among admissions. The difference $C_A-C_R$ measures which Gold class is more likely to receive a score. Overall coverage mixes these two rates according to the cohort's class proportions.

A qualifying item can be lost before scoring or rejected afterwards. Upstream admission loss is $U_A=1-C_A$. Decision loss $D_A(\tau)$ is the fraction of all Gold-Admissible items that receive valid scores below $\tau$. The three outcomes partition the positive denominator:
\begin{equation}
R(\tau)=1-U_A-D_A(\tau)=C_A-D_A(\tau)\leq C_A.
\label{eq:loss}
\end{equation}
The identity identifies where an intervention must act. Changing a threshold changes decision loss among scored items. Recovering a missing score changes upstream loss and exposes that item to the decision rule.

\subsection{Feasible calibration and matched ranking}

For a requested development retention $\rho$, first count valid positive DEV scores. Full-stream feasibility requires
$N_{A,\mathcal{S}}^{\rm DEV}\geq\lceil\rho N_A^{\rm DEV}\rceil$.
We use $\rho=0.9$. Twelve positive DEV items therefore require eleven admissions. If feasible, we choose the largest threshold retaining that integer count, using the corresponding descending order statistic of valid positive scores. Ties are admitted together.

When the target is infeasible, we record that status and apply one conditional fallback: retain at least 90\% of \emph{scoreable} positive DEV items. Its denominator is $N_{A,\mathcal{S}}^{\rm DEV}$ rather than $N_A^{\rm DEV}$. The fallback supplies an operating point under a different target. Invalid scores never enter either ordering, and the selected threshold is fixed for all EVAL cohorts. Implementation details are provided in Appendix~\ref{app:calibration}.

Conditional ROC-AUC and average precision (AP) describe discrimination within $\mathcal{S}$. A comparison on $\mathcal{S}_1\cap\mathcal{S}_2$ holds the evaluated inputs fixed for both score vectors. We report these paired ranking comparisons alongside complete-stream $R/F$ and the loss decomposition. This makes the same evaluation answer both how scores discriminate and what happens to inputs outside the scored population.

\section{Data and Experimental Design}
\label{sec:setup}

\subsection{Human evaluation and the layered resource}

Appendix~\ref{app:annotation} documents the human annotation and adjudication protocol.

The resource contains 3,290 records. Its human-annotated layer contains 872: 602 held-out evaluation items, 100 independent calibration items, and 170 retained development items. Original-302 contains 209 Admissible and 93 Reject items from targeted collection. Random-200 contains 32/168, sampled uniformly after deduplication from a prespecified ten-minute Bluesky window. Challenge-100 contains 31/69, sampled from the same window using semantic-boundary cues for background references, hypotheticals, personal narratives, and opinions. Random tests transfer to an independently sampled live-Web cohort, while Challenge tests difficult admission boundaries.

Independent annotation produced 239/302 initial agreements on Original, with 63 disagreements adjudicated. The combined live-Web package produced 284/300 agreements (94.7\%, $\kappa=0.845$), with all 16 disagreements adjudicated. Evaluation uses the resulting human Gold. DEV30 has 12 Admissible and 18 Reject items within the retained development material. Independent DEV100 and the derived DEV130 union support Qwen calibration sensitivity. Evaluation, development, and reserve allocations are separated by item-ID, normalized-text, external-URL, and near-duplicate checks.

The separate AI-assisted observational layer contains 818 topic-targeted Twitter items (555 Admissible, 263 Reject) and 1,600 random live-Web items (322/1,278). The latter draws 800 from an unused original-window pool and 800 from an independent second window. Two AI annotators agreed on 1,543 items, and humans adjudicated all 57 disagreements. Consensus cases retain AI labels. This layer broadens the resource but supplies neither Gold EVAL nor calibration labels.

\subsection{Strategies, backbones, and paired precision}

We compare three strategies under the same admission policy. \emph{Validity} extracts up to three event-like expressions, scores each with a candidate-validity judgment, and takes the maximum. It requires extraction but not target retrieval. \emph{Direct} applies the policy to the complete item without candidate extraction. The diagnostic \emph{\probe{}} combines 20 extraction, validity/salience, and retrieval features. BM25 \citep{robertson2009bm25} retrieves up to 50 entries from a fixed Wikipedia entity catalog, followed by BGE reranking \citep{chen2024bgem3}.

Extraction returns a JSON array of exact contiguous body spans, each at most 160 characters, within a 256-token generation budget. A valid empty array is a no-candidate outcome, not a formatting failure. Binary judgments use the final-position lowercase \texttt{yes}-minus-\texttt{no} logit. Joint uses StandardScaler and logistic regression, selecting regularization by three-fold DEV30 ROC-AUC. Missing feature components are filled during fitting, but a separate runtime gate excludes incomplete feature vectors from operational scores. Mask-aware threshold selection uses this runtime availability mask while keeping the fitted probe fixed.

The backbones are Qwen2.5-7B \citep{qwen2025technical}, Mistral-7B, Phi-4-mini, GLM-4-9B, Qwen3-14B, and Gemma3-12B in their instruction/chat configurations. The main comparison uses NF4 INT4 weights with BF16 computation, native chat templates, deterministic generation, single-token \texttt{yes}/\texttt{no}, and finite logits. Qwen3 disables thinking. OLMo did not pass compatibility validation and is excluded.

The paired study compares the frozen NF4 INT4 scores with INT8 for Qwen2.5, Mistral, and Phi, plus a Phi BF16 diagnostic. All conditions evaluate the same Original-60 IDs, selected by label-blind hashing, with 45 Admissible and 15 Reject items. Each condition calibrates on DEV30's 12 positive items. Prompt semantics, tokenizer, extraction limits, and score definitions remain fixed within a backbone. Precision conditions are reported together rather than selected by EVAL performance.

\subsection{Mechanism diagnostics and statistical comparisons}

Three extraction-recovery arms independently start from baseline missing-score cases. Retry applies syntax repair and at most one deterministic generation with a format reminder. JSON-schema keeps the baseline task, prompt, model, input, and budget fixed while constraining decoding. Span-safe changes the output interface to source indices, which code slices into exact spans. Baseline valid-empty outputs are excluded from regeneration. Recovered candidates pass through frozen downstream processing and the original historical thresholds. Gold labels and Direct predictions do not guide recovery. These fixed-threshold diagnostics are distinct from the mask-aware main comparison.

Calibration uses DEV data alone. The Qwen DEV30/100/130 sensitivity comparison also fixes EVAL scores, masks, and the fitted Joint probe. Historical predictions and decisions are retained separately from derived mask-aware results. Replay of 54 historical evaluation cells reproduced their decisions. AP uses standard grouped-score handling of ties, correcting a legacy positive-first helper without changing scores, ROC-AUC, thresholds, or fitted probes. Threshold selection uses positive score order statistics, and probe selection uses DEV ROC-AUC.

We use 2,000 paired bootstrap draws, stratified by Gold class and clustered by recorded document groups, with seed 20260928. Each paired AUC contrast resamples identical scored IDs for both methods. Coverage intervals retain the respective Gold-class denominators and masks. Results across configurations share items, so their nominal 95\% intervals are correlated and are not adjusted for multiple comparisons.

\section{Results}
\label{sec:results}

\begin{table*}[t]
\caption{Cross-backbone evaluation on Original-302 (209 Admissible, 93 Reject). $C$, $C_A$, and $C_R$ denote overall and class-conditional coverage. AUC is conditional on each row's scored population. Full: DEV30 full-stream 90\% target feasible. Cond.: target infeasible (10/12 positive scores), conditional fallback applied. Bold/underline mark highest/second retention and lowest/second FAR within each backbone, not a joint winner. All operational rates retain complete-stream class denominators.}
\label{tab:stage7-cross-backbone}
\centering
\small
\begin{tabular}{llrrrrrrl}
\toprule
Model & Method & $C$ & $C_A$ & $C_R$ & AUC$\uparrow$ & Ret.$\uparrow$ & FAR$\downarrow$ & DEV rule \\
\midrule
Qwen2.5-7B & Validity & 0.844 & 0.885 & 0.753 & 0.895 & \underline{0.804} & 0.258 & Full \\
 & Direct & 1.000 & 1.000 & 1.000 & 0.875 & \textbf{0.837} & \underline{0.247} & Full \\
 & Joint & 0.844 & 0.885 & 0.753 & 0.844 & 0.732 & \textbf{0.183} & Full \\
\midrule
Mistral-7B & Validity & 0.821 & 0.861 & 0.731 & 0.898 & \textbf{0.828} & 0.366 & Cond. \\
 & Direct & 0.997 & 0.995 & 1.000 & 0.846 & \underline{0.742} & \textbf{0.269} & Full \\
 & Joint & 0.821 & 0.861 & 0.731 & 0.733 & 0.608 & \underline{0.301} & Cond. \\
\midrule
Phi-4-mini & Validity & 0.699 & 0.780 & 0.516 & 0.873 & \underline{0.718} & \textbf{0.258} & Cond. \\
 & Direct & 1.000 & 1.000 & 1.000 & 0.892 & \textbf{0.990} & 0.774 & Full \\
 & Joint & 0.699 & 0.780 & 0.516 & 0.629 & 0.512 & \underline{0.290} & Cond. \\
\midrule
GLM-4-9B & Validity & 0.944 & 0.981 & 0.860 & 0.819 & \textbf{0.938} & \underline{0.559} & Full \\
 & Direct & 1.000 & 1.000 & 1.000 & 0.759 & \underline{0.885} & 0.602 & Full \\
 & Joint & 0.944 & 0.981 & 0.860 & 0.709 & 0.756 & \textbf{0.473} & Full \\
\midrule
Qwen3-14B & Validity & 0.937 & 0.986 & 0.828 & 0.905 & \underline{0.909} & \underline{0.290} & Full \\
 & Direct & 1.000 & 1.000 & 1.000 & 0.907 & \textbf{0.962} & 0.527 & Full \\
 & Joint & 0.937 & 0.986 & 0.828 & 0.811 & 0.789 & \textbf{0.269} & Full \\
\midrule
Gemma3-12B & Validity & 0.983 & 0.995 & 0.957 & 0.860 & \underline{0.938} & 0.548 & Full \\
 & Direct & 1.000 & 1.000 & 1.000 & 0.917 & \textbf{0.962} & \underline{0.409} & Full \\
 & Joint & 0.980 & 0.995 & 0.946 & 0.757 & 0.656 & \textbf{0.290} & Full \\
\bottomrule
\end{tabular}
\end{table*}

\begin{table*}[t]
\caption{Precision sensitivity on identical Original-60 items (45 Admissible, 15 Reject), distinct from DEV30 (12 Admissible, 18 Reject). Frozen NF4 INT4 is the reference. AUC compares reference/test on their scored intersection. INT4 rows use the INT8 intersection; Phi BF16 uses its own intersection. Each condition is calibrated on DEV30 with the same mask-aware rule. Cond. marks an infeasible full-stream target and conditional fallback. Operating-rate highlights compare precisions within model/method. BF16 is supplementary, not a replacement selected on EVAL. Operating rates are not directly comparable with Table~\ref{tab:stage7-cross-backbone} because the evaluation populations differ (Original-60 versus Original-302).}
\label{tab:stage7-quantization}
\centering
\small
\begin{tabular}{lllrrrrrrl}
\toprule
Model & Precision & Method & $C_A$ & $C_R$ & Shared $n$ & AUC$\uparrow$ (ref./test) & Ret.$\uparrow$ & FAR$\downarrow$ & DEV rule \\
\midrule
Qwen2.5-7B & NF4 INT4 & Validity & 0.822 & 0.800 & 45 & 0.957/0.957 & \underline{0.756} & \textbf{0.200} & Full \\
 & INT8 & Validity & 0.800 & 0.867 & 45 & 0.957/0.962 & \textbf{0.778} & \underline{0.267} & Full \\
 & NF4 INT4 & Direct & 1.000 & 1.000 & 60 & 0.936/0.936 & \underline{0.911} & \textbf{0.267} & Full \\
 & INT8 & Direct & 1.000 & 1.000 & 60 & 0.936/0.910 & \textbf{0.933} & \underline{0.333} & Full \\
\midrule
Mistral-7B & NF4 INT4 & Validity & 0.867 & 0.867 & 47 & 0.905/0.905 & \textbf{0.844} & \underline{0.533} & Cond. \\
 & INT8 & Validity & 0.822 & 0.733 & 47 & 0.905/0.910 & \underline{0.800} & \textbf{0.267} & Full \\
 & NF4 INT4 & Direct & 0.978 & 1.000 & 59 & 0.930/0.930 & \underline{0.756} & \underline{0.133} & Full \\
 & INT8 & Direct & 0.978 & 1.000 & 59 & 0.930/0.975 & \textbf{0.867} & \textbf{0.067} & Full \\
\midrule
Phi-4-mini & NF4 INT4 & Validity & 0.822 & 0.533 & 17 & 0.844/0.844 & \textbf{0.756} & \underline{0.200} & Cond. \\
 & INT8 & Validity & 0.400 & 0.067 & 17 & 0.844/0.813 & \underline{0.400} & \textbf{0.067} & Cond. \\
 & BF16 diag. & Validity & 0.511 & 0.133 & 22 & 0.775/0.900 & \underline{0.400} & \textbf{0.067} & Cond. \\
 & NF4 INT4 & Direct & 1.000 & 1.000 & 60 & 0.928/0.928 & \textbf{1.000} & \underline{0.667} & Full \\
 & INT8 & Direct & 1.000 & 1.000 & 60 & 0.928/0.985 & \textbf{1.000} & \underline{0.667} & Full \\
 & BF16 diag. & Direct & 1.000 & 1.000 & 60 & 0.928/0.970 & \textbf{1.000} & \textbf{0.533} & Full \\
\bottomrule
\end{tabular}
\end{table*}

\begin{table}[t]
\caption{DEV-only operating points. A: DEV30 positive score availability is 12/12 unless marked $10/12$; those four strategies use conditional fallback. B: Qwen fixed-probe threshold sensitivity, with each development set\textquotesingle s full-stream target feasible. Historical artifacts remain unchanged.}
\label{tab:thresholds}
\centering\small
\begin{tabular}{lrrr}
\toprule
\multicolumn{4}{l}{A. Six-backbone DEV30 thresholds}\\
Model & Validity & Direct & Joint \\
\midrule
Qwen2.5-7B & 3.250 & 0.500 & -0.150 \\
Mistral-7B & 9.875$^{10/12}$ & 9.313 & -0.117$^{10/12}$ \\
Phi-4-mini & 3.250$^{10/12}$ & -0.500 & -0.092$^{10/12}$ \\
GLM-4-9B & -0.625 & -1.250 & -0.457 \\
Qwen3-14B & 11.500 & -13.250 & -0.316 \\
Gemma3-12B & 8.750 & 7.500 & 0.448 \\
\midrule
\multicolumn{4}{l}{B. Qwen thresholds with additional calibration data}\\
Method & DEV30 & DEV100 & DEV130\\
\midrule
Validity & 3.250 & -7.500 & -2.375 \\
Direct & 0.500 & -13.250 & -13.000 \\
Joint & -0.150 & -1.655 & -1.007 \\
\bottomrule
\end{tabular}
\end{table}

\begin{table}[t]
\caption{Original-302: $\Delta\mathrm{AUC}$ is Direct minus Validity AUC on identical scored items. Its 95\% intervals use 2,000 paired stratified cluster-bootstrap draws, seed 20260928. A/R give shared Gold-class counts.}
\label{tab:shared}
\centering\small
\begin{tabular}{lrrrr}
\toprule
Model & A/R & Valid./Direct & $\Delta\mathrm{AUC}$ & 95\% CI \\
\midrule
Qwen2.5 & 185/70 & 0.895/0.886 & -0.009 & [-0.050, 0.032] \\
Mistral & 180/68 & 0.898/0.851 & -0.047 & [-0.093, 0.002] \\
Phi-4-mini & 163/48 & 0.873/0.894 & 0.021 & [-0.028, 0.072] \\
GLM-4 & 205/80 & 0.819/0.741 & -0.079 & [-0.146, -0.013] \\
Qwen3 & 206/77 & 0.905/0.898 & -0.008 & [-0.040, 0.024] \\
Gemma3 & 208/89 & 0.860/0.913 & 0.054 & [0.012, 0.096] \\
\bottomrule
\end{tabular}
\end{table}

\subsection{Systems select different evaluation populations}
\label{sec:populations}

Table~\ref{tab:stage7-cross-backbone} shows that the same Original-302 inputs produce different scored populations. Validity scores 211--297 items across the six backbones, compared with 301--302 for Direct. Joint shares most candidate-dependent losses, with an additional feature gate: Gemma scores 296 items under Joint and 297 under Validity. The population entering a ranking metric is therefore determined partly by the system being evaluated.

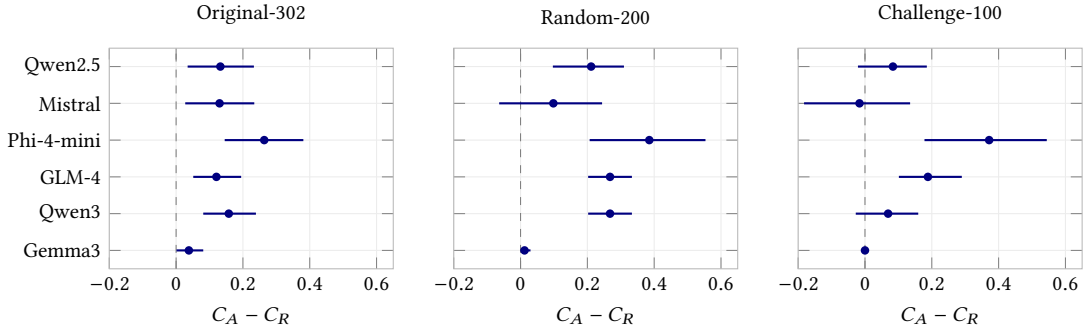
\begin{figure*}[t]
\centering
\begin{tikzpicture}
\begin{groupplot}[group style={group size=3 by 1,horizontal sep=8mm},width=.30\textwidth,height=4.5cm,xmin=-.2,xmax=.65,ymin=.5,ymax=6.5,xtick={-.2,0,.2,.4,.6},ytick={1,2,3,4,5,6},xlabel={$C_A-C_R$},tick label style={font=\small},label style={font=\small},title style={font=\small},grid=major,grid style={gray!15},axis line style={gray!55}]
\nextgroupplot[title={Original-302},yticklabels={Gemma3,Qwen3,GLM-4,Phi-4-mini,Mistral,Qwen2.5}]
\draw[dashed,gray] (axis cs:0,.5)--(axis cs:0,6.5);
\draw[blue!50!black,line width=.8pt] (axis cs:0.034470340073056494,6)--(axis cs:0.23285486443381187,6);
\fill[blue!50!black] (axis cs:0.13247929207182174,6) circle[radius=1.6pt];
\draw[blue!50!black,line width=.8pt] (axis cs:0.027319030714616388,5)--(axis cs:0.2339867263466584,5);
\fill[blue!50!black] (axis cs:0.13006122343983129,5) circle[radius=1.6pt];
\draw[blue!50!black,line width=.8pt] (axis cs:0.14544425580079223,4)--(axis cs:0.3808715336728919,4);
\fill[blue!50!black] (axis cs:0.26377527396203115,4) circle[radius=1.6pt];
\draw[blue!50!black,line width=.8pt] (axis cs:0.05134537222822455,3)--(axis cs:0.19473169727838657,3);
\fill[blue!50!black] (axis cs:0.1206461902556979,3) circle[radius=1.6pt];
\draw[blue!50!black,line width=.8pt] (axis cs:0.08123681638112878,2)--(axis cs:0.23892576014817102,2);
\fill[blue!50!black] (axis cs:0.15768894376704223,2) circle[radius=1.6pt];
\draw[blue!50!black,line width=.8pt] (axis cs:0.0011833101816123293,1)--(axis cs:0.08123681638112878,1);
\fill[blue!50!black] (axis cs:0.03822606369295667,1) circle[radius=1.6pt];
\nextgroupplot[title={Random-200},yticklabels={}]
\draw[dashed,gray] (axis cs:0,.5)--(axis cs:0,6.5);
\draw[blue!50!black,line width=.8pt] (axis cs:0.09672619047619047,6)--(axis cs:0.30952380952380953,6);
\fill[blue!50!black] (axis cs:0.21130952380952384,6) circle[radius=1.6pt];
\draw[blue!50!black,line width=.8pt] (axis cs:-0.06398809523809523,5)--(axis cs:0.24404761904761907,5);
\fill[blue!50!black] (axis cs:0.0982142857142857,5) circle[radius=1.6pt];
\draw[blue!50!black,line width=.8pt] (axis cs:0.20684523809523808,4)--(axis cs:0.5535714285714286,4);
\fill[blue!50!black] (axis cs:0.3854166666666667,4) circle[radius=1.6pt];
\draw[blue!50!black,line width=.8pt] (axis cs:0.20238095238095233,3)--(axis cs:0.33333333333333337,3);
\fill[blue!50!black] (axis cs:0.2678571428571429,3) circle[radius=1.6pt];
\draw[blue!50!black,line width=.8pt] (axis cs:0.20238095238095233,2)--(axis cs:0.33333333333333337,2);
\fill[blue!50!black] (axis cs:0.2678571428571429,2) circle[radius=1.6pt];
\draw[blue!50!black,line width=.8pt] (axis cs:0.0,1)--(axis cs:0.029761904761904767,1);
\fill[blue!50!black] (axis cs:0.011904761904761862,1) circle[radius=1.6pt];
\nextgroupplot[title={Challenge-100},yticklabels={}]
\draw[dashed,gray] (axis cs:0,.5)--(axis cs:0,6.5);
\draw[blue!50!black,line width=.8pt] (axis cs:-0.02103786816269293,6)--(axis cs:0.18513323983169705,6);
\fill[blue!50!black] (axis cs:0.0836839644693782,6) circle[radius=1.6pt];
\draw[blue!50!black,line width=.8pt] (axis cs:-0.18232819074333806,5)--(axis cs:0.1351098644226274,5);
\fill[blue!50!black] (axis cs:-0.016362786348761094,5) circle[radius=1.6pt];
\draw[blue!50!black,line width=.8pt] (axis cs:0.17812061711079946,4)--(axis cs:0.5441795231416549,4);
\fill[blue!50!black] (axis cs:0.3716690042075736,4) circle[radius=1.6pt];
\draw[blue!50!black,line width=.8pt] (axis cs:0.10144927536231885,3)--(axis cs:0.28985507246376807,3);
\fill[blue!50!black] (axis cs:0.18840579710144922,3) circle[radius=1.6pt];
\draw[blue!50!black,line width=.8pt] (axis cs:-0.02758298270219728,2)--(axis cs:0.1594202898550725,2);
\fill[blue!50!black] (axis cs:0.06919121084618984,2) circle[radius=1.6pt];
\draw[blue!50!black,line width=.8pt] (axis cs:0.0,1)--(axis cs:0.0,1);
\fill[blue!50!black] (axis cs:0.0,1) circle[radius=1.6pt];
\end{groupplot}
\end{tikzpicture}
\caption{Validity's class-conditional score coverage difference, $C_A-C_R$, with nominal 95\% bootstrap intervals. Positive values indicate a larger fraction of Admissible than Reject inputs receiving scores. The common axes show variation across backbones and cohorts, including Mistral's reverse-direction Challenge result.}
\Description{Three cohort panels show estimates and intervals for Admissible minus Reject coverage across six backbones, with a zero reference line.}
\label{fig:class-coverage}
\end{figure*}

\paragraph{The missing items have a class composition.}
Figure~\ref{fig:class-coverage} shows that overall coverage hides unequal score availability across Gold classes. On Original-302, all six Validity backbones have $C_A>C_R$. Phi's difference is 0.264 (95\% CI [0.145, 0.381]), and Qwen3's is 0.158 [0.081, 0.239]. Across 54 model--strategy--cohort cells, 19 have complete coverage. Of the 35 with missing scores, 32 have $C_A>C_R$ and 25 have nominal intervals entirely above zero. The reverse cases are Mistral Direct on Original and Mistral Validity/Joint on Challenge, none with an interval entirely below zero. The recurring selection favors score availability for Admissible inputs, with observed exceptions.

This distinction matters when collection regimes change. Qwen Validity's overall coverage falls from 84.4\% on Original to 76.0\% on Random, while its Admissible coverage rises from 88.5\% to 93.8\%. Phi's overall coverage falls from 69.9\% to 39.5\%, with Reject coverage declining from 51.6\% to 33.3\%. Gemma instead scores every Challenge item under all three strategies. The overall rate alone cannot distinguish changes in class composition from changes in the scoring process.

\paragraph{Matching inputs changes the ranking comparison.}
On Qwen Original, Validity's AUC of 0.895 uses 255 items, whereas Direct's 0.875 uses all 302. Restricting Direct to the same 255 inputs raises its AUC to 0.886. The apparent gap narrows from 0.020 to 0.009 with neither score vector changed. Table~\ref{tab:shared} extends this comparison to six backbones. On their respective shared subsets, GLM favors Validity ($\Delta=-0.079$, 95\% CI [$-0.146,-0.013$]), while Gemma favors Direct ($\Delta=0.054$, [0.012, 0.096]); $\Delta$ is Direct minus Validity. The other four intervals include zero. Matched comparisons expose backbone-specific ranking differences instead of conflating discrimination with which inputs reached scoring.

Cross-backbone comparisons show the same population sensitivity. Original Validity's six-way scored intersection contains 162 items. Across 135 cross-model pairwise contrasts over methods and cohorts, 15 ordering signs differ between own-population and common-population AUC. These are descriptive changes in ordering, not evidence of a general ranking reversal. Together, the analyses show why adding coverage to an AUC table is useful but incomplete: the identity and Gold-class composition of the scored inputs also matter.

\subsection{Availability limits the operating point, not just the metric}
\label{sec:operation}

The four Mistral/Phi Validity and Joint configurations score ten of twelve positive DEV30 items. Full-stream 90\% retention requires eleven, so the target is infeasible before threshold selection. The other fourteen configurations score all twelve. Table~\ref{tab:thresholds} records the selected operating points. For the four infeasible cases, the conditional target requires nine admissions from the ten scoreable positives, rather than eleven from the full twelve.

The historical implementation obscured this distinction. Unavailable Validity scores entered DEV ordering as finite $-20$ placeholders. Joint similarly ordered margins from filled fitting vectors before enforcing its runtime gate. Mask-aware selection removes these unavailable scores and exposes the infeasible target. With predictions unchanged, Mistral Validity moves from historical $\tau=-20$ to conditional $\tau=9.875$. Original FAR decreases from 68/93 (73.1\%) to 34/93 (36.6\%), and retention changes from 180/209 to 173/209. Phi Validity moves from $-20$ to 3.25, changing FAR from 48/93 to 24/93 and retention from 163/209 to 150/209. Coverage and conditional AUC stay fixed. Correcting threshold selection changes decisions, while the availability bound remains.

\begin{figure}[t]
\centering
\includegraphics[width=\columnwidth]{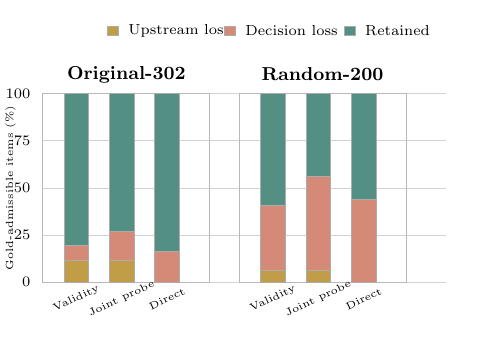}
\caption{Qwen admission loss on Original-302 and Random-200. Upstream score loss, scored-but-rejected decision loss, and retention partition the full Gold-Admissible denominator. Operating points use DEV30 mask-aware thresholds.}
\Description{Stacked bars separate upstream loss, decision loss, and retention for Validity, Joint, and Direct in two evaluation cohorts.}
\label{fig:loss}
\end{figure}

\paragraph{The loss decomposition locates the bottleneck.}
Figure~\ref{fig:loss} separates the two stages of positive loss. Qwen Original Validity loses 24/209 Admissible items before scoring and rejects 17/209 afterwards, retaining 168/209. Joint has the same upstream loss but rejects 32/209 scored positives, retaining 153/209. Direct scores all positives and rejects 34/209, retaining 175/209. On Random, Validity and Joint each lose two of 32 positives upstream but retain 19 and 14, respectively, while Direct retains 18. The same coverage can support different decisions, and the same retention can hide different places to intervene.

\paragraph{A feasible target still leaves a threshold choice.}
Table~\ref{tab:thresholds}B fixes Qwen's fitted probe, EVAL scores, and masks while changing the calibration set. DEV100 has 15/16 scoreable positives under Validity/Joint, and DEV130 has 27/28, meeting their full-stream requirements of 15 and 26 admissions. Validity's DEV30/100/130 thresholds are 3.25, $-7.5$, and $-2.375$. On the same Original inputs, retention is 168/209, 182/209, and 177/209, while FAR is 24/93, 54/93, and 42/93.

Direct isolates threshold sensitivity without upstream score loss. Changing Qwen's calibration set from DEV30 to DEV100 moves $\tau$ from 0.5 to $-13.25$, raising Original retention from 175/209 to 205/209 and FAR from 23/93 (24.7\%) to 85/93 (91.4\%). Scores, masks, and ranking are identical. Thus availability determines what retention is possible, while development-selected thresholds determine which tradeoff is realized and transferred to EVAL.

\begin{figure*}[t]
\centering
\begin{tikzpicture}
\begin{groupplot}[group style={group size=3 by 1,horizontal sep=9mm},width=.30\textwidth,height=4.2cm,xmin=0,xmax=60,ymin=.4,ymax=2.6,xtick={0,15,30,45,60},ytick={1,2},yticklabels={Direct,Validity},xlabel={Items},tick label style={font=\small},label style={font=\small},title style={font=\small},axis line style={gray!55},legend style={draw=none,font=\small},legend columns=4]
\nextgroupplot[title={Qwen2.5-7B},legend to name=masklegend]
\fill[teal!55!gray] (axis cs:0,1.77) rectangle (axis cs:45,2.23);
\fill[orange!60!gray] (axis cs:45,1.77) rectangle (axis cs:49,2.23);
\fill[blue!50!gray] (axis cs:49,1.77) rectangle (axis cs:53,2.23);
\fill[gray!20] (axis cs:53,1.77) rectangle (axis cs:60,2.23);
\fill[teal!55!gray] (axis cs:0,0.77) rectangle (axis cs:60,1.23);
\addlegendimage{area legend,fill=teal!55!gray,draw=none}
\addlegendentry{Scored in both}
\addlegendimage{area legend,fill=orange!60!gray,draw=none}
\addlegendentry{Lost score}
\addlegendimage{area legend,fill=blue!50!gray,draw=none}
\addlegendentry{Gained score}
\addlegendimage{area legend,fill=gray!20,draw=none}
\addlegendentry{Missing in both}
\nextgroupplot[title={Mistral-7B}]
\fill[teal!55!gray] (axis cs:0,1.77) rectangle (axis cs:47,2.23);
\fill[orange!60!gray] (axis cs:47,1.77) rectangle (axis cs:52,2.23);
\fill[blue!50!gray] (axis cs:52,1.77) rectangle (axis cs:53,2.23);
\fill[gray!20] (axis cs:53,1.77) rectangle (axis cs:60,2.23);
\fill[teal!55!gray] (axis cs:0,0.77) rectangle (axis cs:59,1.23);
\fill[gray!20] (axis cs:59,0.77) rectangle (axis cs:60,1.23);
\nextgroupplot[title={Phi-4-mini}]
\fill[teal!55!gray] (axis cs:0,1.77) rectangle (axis cs:17,2.23);
\fill[orange!60!gray] (axis cs:17,1.77) rectangle (axis cs:45,2.23);
\fill[blue!50!gray] (axis cs:45,1.77) rectangle (axis cs:47,2.23);
\fill[gray!20] (axis cs:47,1.77) rectangle (axis cs:60,2.23);
\fill[teal!55!gray] (axis cs:0,0.77) rectangle (axis cs:60,1.23);
\end{groupplot}
\end{tikzpicture}
\par\ref{masklegend}
\caption{Score-availability transitions from NF4 INT4 to INT8 on the same Original-60 inputs. Four states account for every item: scored in both, lost, gained, and missing in both. Qwen Validity exchanges four items despite unchanged coverage, while Direct's masks remain stable.}
\Description{Three backbone panels show Validity and Direct masks as four colored segments for scored in both conditions, lost scores, gained scores, and missing in both.}
\label{fig:mask-transitions}
\end{figure*}
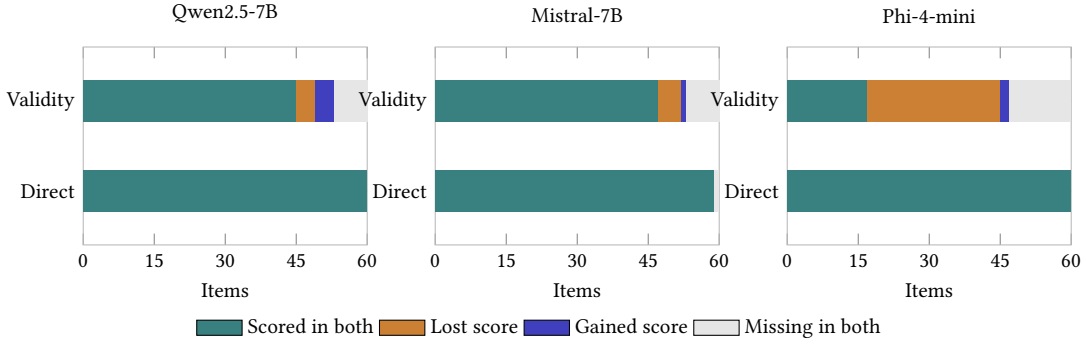

\subsection{Execution precision changes who reaches the decision rule}
\label{sec:precision}

The cross-backbone comparison changes the system. Table~\ref{tab:stage7-quantization} tests the same dependency while holding backbone and input IDs fixed. NF4 INT4 supplies the reference, INT8 the paired perturbation, and Phi BF16 an additional diagnostic. Original-60 measures evaluation outcomes, while DEV30 supplies a separate calibration population.

\paragraph{An unchanged coverage rate can hide different inputs.}
Qwen Validity scores 49/60 items under both INT4 and INT8. Four previously scored inputs lose scores, while four previously unscored inputs gain them, leaving an intersection of 45. Admissible/Reject counts change from 37/45 and 12/15 to 36/45 and 13/15. Figure~\ref{fig:mask-transitions} shows the corresponding transitions for all three backbones. Mistral loses five and gains one, and Phi loses 28 and gains two. Direct preserves its scoreable IDs: all 60 for Qwen and Phi, and the same 59 for Mistral. Paired masks reveal changes that a net coverage difference cannot describe.

\paragraph{DEV feasibility and EVAL coverage can move differently.}
Mistral Validity scores 10/12 positive DEV items under INT4 and 12/12 under INT8, making the full-stream target feasible. Yet Original-60 coverage decreases from 52/60 to 48/60, and Admissible coverage from 39/45 to 37/45. At each condition's DEV-selected threshold, retention changes from 38/45 to 36/45 and FAR from 8/15 to 4/15. Feasibility on the development population and coverage on the evaluation population capture different consequences of the same configuration change.

Phi's Admissible EVAL coverage falls from 37/45 under INT4 to 18/45 under INT8, with BF16 scoring 23/45. Positive DEV counts are 10/12, 3/12, and 4/12, so all remain infeasible. Higher numerical precision therefore does not consistently improve score availability. The observed masks establish this execution sensitivity, but do not identify the extraction or validation cause of each lost score.

\paragraph{Stable masks leave room for ranking changes.}
Mistral Direct's AUC rises from 0.930 to 0.975 on its unchanged shared subset ($\Delta=0.045$, 95\% CI [0.008, 0.098]). Qwen Direct changes from 0.936 to 0.910 ($\Delta=-0.026$, [$-0.076$, 0.016]). Stable availability therefore does not make predictions precision-invariant. Phi Validity's shared intersections are much smaller: 16 Admissible and one Reject for INT4--INT8, and 20 and two for INT4--BF16. Their AUCs in Table~\ref{tab:stage7-quantization} describe these small intersections rather than a stable precision ranking. The paired evidence distinguishes changes in scoring access from changes in discrimination and the resulting operating point.

\subsection{Recovering scores exposes the downstream tradeoff}
\label{sec:recovery}

The availability changes raise a mechanism question: how much missingness can extraction repair remove? Inde\-pendent Retry, JSON-schema, and Span-safe diagnostics test this directly. JSON-schema recovers 4, 0, and 0 Qwen scores on Original, Random, and Challenge, compared with 11, 13, and 6 for Mistral. Retry recovers 14/6/2 and 36/30/11, respectively. Format constraints and an additional generation opportunity therefore have different effects even before the stronger span-index interface is introduced.

Span-safe recovers 44/54, 45/54, and 14/15 of Mistral's missing scores (81.5--93.3\%). Original has 53 extraction-eligible cases among its 54 missing scores. Qwen has 104 missing scores across the cohorts: 60 baseline valid-empty cases are excluded by protocol, and all 44 actual Span-safe attempts return valid-empty structures without parse, interval, range, or recorded token-budget failure. This residual is not an inability to follow the index-output protocol. Phi Original has 91 missing scores, skips 75 baseline valid-empty cases, and recovers 3, 2, and 15 under Retry, JSON-schema, and Span-safe. Recoverability thus depends on both the system and the intervention, with persistent valid-empty outputs alongside recoverable failures.

Recovery also changes which inputs reach a fixed threshold. Under Mistral's historical $\tau=-20$, which admits every scoreable input, Original Span-safe raises coverage from 82.1\% to 96.7\%, retention from 86.1\% to 98.6\%, and FAR from 73.1\% to 92.5\%. More qualifying items enter, but so do more Gold-Reject items. These historical-point diagnostics hold the boundary fixed, whereas Table~\ref{tab:stage7-cross-backbone} uses mask-aware calibration. Their operational lesson is that upstream recovery expands the decisions available to a system without choosing an appropriate decision boundary for the expanded population.

\section{Conclusion}
\label{sec:conclusion}

Open-Web event admission begins with a complete item, not a successful extraction or retrieved target. Evaluating that item throughout the pipeline reveals a system property hidden by scored-only ranking: processing determines who can receive a decision score. Across six backbones, scored populations differ in both size and class composition. Precision perturbations show that this selection can change even when the backbone, inputs, and aggregate coverage remain fixed.

The consequences reach beyond the ranking comparison. Score availability bounds attainable retention, while the selected threshold governs decisions among scoreable inputs. Recovery and calibration sensitivity show why improving upstream access and improving complete-stream operation are separate achievements. Our evaluation connects them through matched ranking populations, a DEV feasibility check, and a decomposition of upstream and decision losses. Overall coverage becomes a diagnostic quantity rather than a standalone quality score.

The evidence spans three English-language event-admission cohorts and a smaller paired precision study. Broader collection periods, languages, and downstream clustering or timeline utility are directions for extending this evaluation. For multistage Web systems, the immediate implication is concrete: judge the pipeline on the inputs it receives, and make visible both the inputs it cannot score and the decisions it makes about the rest.

\section*{Data and Code Availability}
The public archive location for human annotations, annotation protocols, frozen predictions, and evaluation resources will be provided in a subsequent version. Item-level scores and availability masks support recalculation of calibration thresholds, ranking metrics, complete-stream counts, and uncertainty estimates. Recovery records retain intervention outcomes separately from the main mask-aware evaluation.

\bibliographystyle{ACM-Reference-Format}
\bibliography{references}
\appendix
\section{Annotation Protocol and Adjudication}
\label{app:annotation}

\paragraph{Annotation preparation.}
Multiple pilot rounds and guideline discussions aligned interpretations of event elements, granularity, and the admission boundary. Annotators judged whether the complete item foregrounded concrete event information suitable for organization, rather than merely mentioning a linkable target. Actions, decisions, changes, outcomes, and personal occurrences can qualify without complete dates, locations, or knowledge-base entries. Wishes, hypotheticals, and questions do not by themselves establish an occurrence.

\paragraph{Displayed evidence and independent labels.}
Two researchers independently assigned \admissible{} or \reject{} using the displayed body, timestamp, and reply, quote, and media flags. Links and unseen parent posts or media were not consulted. Labels concern displayed information, not externally verified truth. Live-Web packages used independently shuffled orders and hid peer answers, model predictions, and challenge cues. AI assistance in event-element preparation did not determine Gold labels.

\paragraph{Human adjudication.}
Disagreements received human adjudication. Original's 200-item extension and Live-Web-300 used blinded adjudication packages. Original's earlier 102-item component used controlled human adjudication, not a third independent blind annotation. Eight follow-up judgments used machine translation, with final choices made by a human and no automatic labeling. Raw A/B labels remain separate from final decisions. These 602 evaluation items are distinct from the AI-assisted observational layer.

\paragraph{Inter-annotator agreement.}
Original's component agreements were 85/102 and 154/200: 239/302 overall (79.1\%), with all 63 disagreements (20.9\%) adjudicated. Live-Web-300, comprising Random-200 and Challenge-100, had 284/300 agreements (94.7\%) and 16 disagreements (5.3\%). A/B assigned 58/74 positive labels, yielding Cohen's $\kappa=0.845$. All 16 disagreements were adjudicated: five \admissible{} and eleven \reject{}. Collection regimes and item distributions differ, so agreement differences are not attributed to training alone. No pooled Original $\kappa$ is reported.

\paragraph{Traceable boundary cases.}
These anonymized paraphrases are disagreements from Original's 200-item extension. Labels are unchanged. Explanations are retrospective unless explicitly attributed to a recorded adjudication reason.

\begin{itemize}
\item \textbf{Personal occurrence.} Record \texttt{S4-FH200-0111} reports a flight delay, tentatively attributing it to an IT outage, alongside personal commentary. A: \reject{}; B: \admissible{}; final: \admissible{}. No reason was recorded. Retrospectively, the delay supplies a concrete occurrence without requiring verification of its proposed cause.
\item \textbf{Predicted developments.} Record \texttt{S4-FH200-0084} forecasts future headset applications, lower prices, and adoption. A: \admissible{}; B: \reject{}; final: \reject{}. No reason was recorded. Retrospectively, a named product and dated predictions do not establish a reported occurrence.
\item \textbf{Commentary and advice.} Record \texttt{S4-FH200-0083} evaluates a headset, recommends a store demonstration, and advises against purchase. A: \reject{}; B: \admissible{}; final: \reject{}. The recorded reason, translated from Chinese, identifies commentary without an event anchor, distinguishing product references from foregrounded event information.
\end{itemize}
\section{Mask-Aware Calibration and Reproducibility Details}
\label{app:calibration}

\paragraph{Threshold selection.}
Inputs are DEV labels $y_i$, scores $s_i$, and availability masks $c_i$. The implementation fixes $\rho=0.9$ and requires at least one Gold-Admissible DEV item.

\begin{quote}
\textbf{Mask-aware threshold selection}\par
\begin{tabbing}
\quad\=\quad\=\kill
$N_A\leftarrow\sum_i y_i$; $k_{\mathrm{full}}\leftarrow\lceil\rho N_A\rceil$.\\
$P\leftarrow\{i:y_i=1\text{ and }c_i=1\}$.\\
Sort $P$ by decreasing $s_i$, then by item ID.\\
$m\leftarrow |P|$; $f\leftarrow(m\geq k_{\mathrm{full}})$.\\
\textbf{if} $m=0$: \textbf{return} $(\mathrm{None},\mathrm{false},\text{conditional})$.\\
\textbf{if} $f$: $k\leftarrow k_{\mathrm{full}}$; rule $\leftarrow$ full-stream.\\
\textbf{else}: $k\leftarrow\lceil\rho m\rceil$; rule $\leftarrow$ conditional.\\
$\tau\leftarrow s_{P[k]}$ \quad (one-based index).\\
\textbf{return} $(\tau,f,\text{rule})$.
\end{tabbing}
\end{quote}

The tuple summarizes the returned record's threshold, feasibility, and rule. Even when $m=0$, the record retains the conditional-fallback flag, but no threshold is available.

\paragraph{Availability and boundary behavior.}
Input preparation combines strategy-specific runtime validity checks with finite-score checks. Invalid scores and historical placeholders are excluded by the mask before ordering; finiteness alone does not establish availability. Item ID breaks sorting ties reproducibly, but admission uses $s_i\geq\tau$, so all scoreable items tied at the threshold are admitted. The retained count can therefore exceed $k$. If no valid positive DEV score exists, the implementation returns no threshold and reports EVAL retention, FAR, and decision loss as unavailable, rather than supplying a numerical fallback. A DEV set with no Gold positives is outside the implemented routine's supported input domain.

Each backbone, strategy, and execution configuration supplies its own masks and DEV threshold. The selected threshold is fixed across its EVAL cohorts. An unscored item never enters threshold comparison but remains in the complete-stream class denominator. Conditional fallback uses the scoreable positive DEV count $m$, not $N_A$, and does not satisfy an infeasible full-stream target.

\paragraph{Ranking populations.}
Conditional AUC uses each configuration's own scored set $\mathcal{S}$. Paired shared-subset AUC uses the same item IDs in $\mathcal{S}_1\cap\mathcal{S}_2$ for both score vectors, including items below their decision thresholds. These populations are formed from availability masks, not admission decisions. AUC is unavailable when the evaluated scored set lacks either Gold class. Complete-stream operating rates retain the original Gold-class denominators regardless of the ranking subset.
\end{document}